# Phase-field modeling and electronic structural analysis of flexoelectric effect at 180 ° domain walls in ferroelectric PbTiO$_3$

Yu-Jia Wang[1], Jiangyu Li[2,3,a)], Yin-Lian Zhu[1], Xiu-Liang Ma[1,4,b)]

[1]Shenyang National Laboratory for Materials Science, Institute of Metal Research, Chinese Academy of Sciences, 72 Wenhua Road, Shenyang, Liaoning, 110016, China

[2]Shenzhen Key Laboratory of Nanobiomechanics, Shenzhen Institutes of Advanced Technology, Chinese Academy of Sciences, University Town of Shenzhen, Shenzhen, Guangdong, 518055, China

[3]Department of Mechanical Engineering, University of Washington, Seattle, Washington 98195-2600, USA

[4]School of Materials Science and Engineering, Lanzhou University of Technology, Langongping Road 287, 730050 Lanzhou, China

Flexoelectric effect is the coupling between strain, polarization and their gradients, which are prominent at the nanoscale. Although this effect is important to understand nanostructures, such as domain walls in ferroelectrics, its electronic mechanism is not clear. In this work, we combined phase-field simulations and first-principles calculations to study the 180 ° domain walls in tetragonal ferroelectric PbTiO$_3$, and found that the ultimate source of Néel components is the gradient of the square of spontaneous polarizations. Electronic structural analysis reveals that there is a redistribution of electronic charge density and potential around domain walls, which produces the electric field and Néel components. This work thus sheds light on the electronic mechanism of the flexoelectric effect around 180 ° domain walls in tetragonal ferroelectrics.

# I. INTRODUCTION

Flexoelectricity describes the class of physical phenomena, that strain (stress) gradient induces electric polarization (field), or electric polarization (field) gradient results in strain (stress). Recently, many studies investigated the effect of flexoelectricity on materials' properties, especially in ferroelectrics.[1-15] Abundant interfaces exist in ferroelectric materials, such as domain walls (DWs) and morphotropic phase boundaries (MPBs), which provide natural places where gradients in polarization and/or strain (stress) arise. Thus, flexoelectric effect should be prominent at these interfaces. Indeed, many interesting properties of ferroelectric DWs or MPBs are found to be related to the flexoelectric effect. For example, Catalan et al. found that there exists a strain distribution around the 90°DW near the substrate in PbTiO$_3$ (PTO) films, which causes polarization rotations through the flexoelectric effect.[3] The normally uncharged DWs in BiFeO$_3$ and Pb(Zr,Ti)O$_3$ are found to be conductive,[16, 17] which may be caused by the Néel-type polarization components due to the flexoelectric effect.[18] The MPB between R-like and T-like BiFeO$_3$ shows many interesting properties, such as high piezoelectric and magnetic responses[19] and a large enhancement in the anisotropic interfacial photocurrent,[12] which may be explained by considering the flexoelectric effect as well.[20]

180°DWs in tetragonal ferroelectrics are generally 1~2 nm thick, near which large polarization and strain gradients exist. Thus the flexoelectric effect is very important to understand the structure of 180°DWs. The main feature of a 180°DW is its tangential polarization profile: The polarization vectors rapidly shrink their magnitudes near DWs and reverse their directions upon crossing DW planes. Accompanying such polarization profile, there is also a strain distribution due to the electrostrictive effect. This is the classical Ising-type 180°DWs. Recent studies, however, showed that the structures of 180°DWs are far more complex. Besides the Ising-type components, that other components of Bloch- and Néel-type could also exist around 180°DWs.[21-28] Phenomenological analysis and phase-field simulations indicate that the flexoelectric effect should be taken into account to understand the emergence of these non-Ising characters.[26-28] However, it is not clear how these components actually emerge, especially at the electronic level.

In this paper, with the aim to understand the importance of flexoelectricity on 180° DWs and its electronic mechanisms, we first used phase-field simulations to study the effects of flexoelectric coefficients on the structure of 180° DWs and then used first-principles calculations to further understand the electronic origin of flexoelectric effect, taking the [100]-oriented 180° DW in PTO as an example. We found that the ultimate source of Néel components is the gradient of the square of spontaneous polarization. Electronic structural analysis reveals that a redistribution of electronic charge density and potential is induced around DWs, which produces the electric field between the DW region and the bulk region and finally induces Néel components.

## II. CALCULATION METHODS

### A. PHASE-FIELD MODEL

3D phase-field models are developed to study the effect of flexoelectricity on the polarization and strain characteristics around DWs in tetragonal ferroelectrics. The order parameters are chosen as the three components of polarization vectors. The system energy is the functional of polarization and strain:

$$F = F_{bulk} + F_{grad} + F_{elas} + F_{elec} \quad (1)$$

The first term is the bulk energy or Landau-Devonshire energy:

$$\begin{aligned}F_{bulk} = {} & \alpha_1(P_1^2 + P_2^2 + P_3^2) + \alpha_{11}(P_1^4 + P_2^4 + P_3^4) \\ & + \alpha_{12}(P_1^2 P_2^2 + P_2^2 P_3^2 + P_3^2 P_1^2) + \alpha_{111}(P_1^6 + P_2^6 + P_3^6) \\ & + \alpha_{112}[P_1^4(P_2^2 + P_3^2) + P_2^4(P_3^2 + P_1^2) + P_3^4(P_1^2 + P_2^2)] + \alpha_{123} P_1^2 P_2^2 P_3^2\end{aligned} \quad (2)$$

which could be used to describe a first order ferroelectric phase transition.

The second term is the gradient energy or Ginzburg energy, whose contribution to the total system is reflected as the DW energy:

$$F_{grad} = \frac{1}{2} G_{11}(P_{1,1}^2 + P_{2,2}^2 + P_{3,3}^2) + G_{12}(P_{1,1}P_{2,2} + P_{2,2}P_{3,3} + P_{3,3}P_{1,1})$$
$$+ \frac{1}{2} G_{44}\left[(P_{1,2} + P_{2,1})^2 + (P_{2,3} + P_{3,2})^2 + (P_{3,1} + P_{1,3})^2\right] \quad (3)$$

The third term is the elastic energy:

$$F_{elas} = \frac{1}{2} C_{ijkl} e_{ij} e_{kl} = \frac{1}{2} C_{ijkl}(\varepsilon_{ij} - \varepsilon_{ij}^o - \varepsilon_{ij}^f)(\varepsilon_{kl} - \varepsilon_{kl}^o - \varepsilon_{kl}^f) \quad (4)$$

where $C_{ijkl}$ is the elastic stiffness tensor and $e_{ij}$, $\varepsilon_{ij}$, $\varepsilon_{ij}^o$, $\varepsilon_{ij}^f$ are the elastic strain, the total strain, the electrostrictive strain, and the flexoelectric strain. The electrostrictive and flexoelectric strains can be calculated by these equations:

$$\varepsilon_{ij}^o = Q_{ijkl} P_k P_l \quad (5a)$$

$$\varepsilon_{ij}^f = -F_{ijkl} P_{k,l} \quad (5b)$$

where $Q_{ijkl}$ and $F_{ijkl}$ are the electrostrictive and flexoelectric coefficients. In Eq. (4), we incorporate the electrostrictive and flexoelectric effects into the elastic energy. As demonstrated in the Supplementary Materials, this expression is the Legendre transformation of the corresponding terms in the elastic Gibbs energy.

The fourth term is the electrostatic energy:

$$F_{elec} = -\frac{1}{2} E_i^{dep} P_i \quad (6)$$

where $E_i^{dep}$ is the depolarization field.

By differentiating these energies with respect to the polarization, we can get the forces to drive the evolution of polarization. Among these forces the mechanical driving force must be handled carefully and the result is:

$$-\frac{\delta F_{elas}}{\delta P_i} = 2 q_{ijkl} P_j e_{kl} + f_{ijkl} e_{kl,j} \quad (7)$$

where $q_{ijkl} = C_{ijmn} Q_{mnkl}$ and $f_{ijkl} = C_{ijmn} F_{mnkl}$ are other types of electrostrictive and flexoelectric coefficients. From this expression, we can see that both the elastic strains and their gradients contribute to the mechanical driving force. The derivation can be found in the Supplementary Materials. Since this mechanical driving force is effectively an electric field, the two terms in this equation can be considered as the electrostriction-induced and flexoelectricity-induced electric fields. The core in Eq. (7) is the elastic strain, which can be obtained by solving the mechanical equilibrium equation (Eq. (8a)). The depolarization field can be found by solving the Poisson's equation (Eq. (8b)). After all the driving forces are found, the Ginzburg-Landau equation (Eq. (8c)) was solved to update the polarization vectors.

$$\sigma_{ij,j} = 0 \tag{8a}$$

$$D_{i,i} = 0 \tag{8b}$$

$$\frac{\partial P_i}{\partial t} = -L \frac{\delta F}{\delta P_i} \tag{8c}$$

where $D$ is the electric displacement and $L$ is the dynamical parameter.

TABLE I. Material parameters of $PbTiO_3$[29]

| | | |
|---|---|---|
| $\alpha_1 = -1.725 \times 10^8$ C$^{-2}$ m$^2$ N | $\alpha_{11} = -7.3 \times 10^7$ C$^{-4}$ m$^6$ N | $\alpha_{12} = 7.5 \times 10^8$ C$^{-4}$ m$^6$ N |
| $\alpha_{111} = 2.6 \times 10^8$ C$^{-6}$ m$^{10}$ N | $\alpha_{112} = 6.1 \times 10^8$ C$^{-6}$ m$^{10}$ N | $\alpha_{123} = -3.7 \times 10^8$ C$^{-6}$ m$^{10}$ N |
| $G_{11} = 4.8 \times 10^{-11}$ C$^{-2}$ m$^4$ N | $G_{12} = -4.8 \times 10^{-11}$ C$^{-2}$ m$^4$ N | $G_{44} = 4.8 \times 10^{-11}$ C$^{-2}$ m$^4$ N |
| $C_{11} = 1.75 \times 10^{11}$ N m$^{-2}$ | $C_{11} = 1.75 \times 10^{11}$ N m$^{-2}$ | $C_{11} = 1.75 \times 10^{11}$ N m$^{-2}$ |
| $Q_{11} = 0.089$ C$^{-2}$ m$^4$ | $Q_{12} = -0.026$ C$^{-2}$ m$^4$ | $Q_{44} = 0.038$ C$^{-2}$ m$^4$ |
| $\alpha_0 = 1.725 \times 10^8$ C$^{-2}$ m$^2$ N | $G_{110} = 6.9 \times 10^{-12}$ C$^{-2}$ m$^4$ N | $P_0 = 0.75$ C m$^{-2}$ |

A grid of $64 \times 2 \times 2$ is used to simulate the 180°DW model, where each grid point corresponds to 0.2 nm. We have tested larger cells ($128 \times 2 \times 2$ and $256 \times 2 \times 2$) and found that the results are consistent with the smaller one. The normal of 180°DWs is parallel to the $x$ direction. In the initial structure, the polarizations of one half of grid points are set along the $+z$ direction, while the other half along the $-z$ direction. The three-dimensional periodic boundary condition is applied and the technique of fast Fourier transformation (FFT) is adopted to solve Eqs. (8a-c) which are second-order partial differentiation equations since FFT could effectively convert the differentiation into the multiplication. By solving the mechanical equilibrium equation, we found that $\sigma_{11}$, $\sigma_{12}$, $\sigma_{13}$, and $\sigma_{23}$ are all zero, while $\sigma_{22}$ and $\sigma_{33}$ are non-zero. To make our simulation accord with the stress-free boundary condition ($\sigma_{ij}|_{x_1 \to \pm\infty} = 0$), constant stresses $\sigma_{22}^0$ and $\sigma_{33}^0$ have be applied to the system to make sure that $\sigma_{22}$ and $\sigma_{33}$ are zero in the grid points away from DWs. The simulation was considered to be convergent when the averaged polarization difference between two sequential simulation steps is less than $1 \times 10^{-8}$. The material parameters are chosen for ferroelectric PTO from the previous literature.[29] The gradient coefficients are chosen to make the DW width the same as the first-principles result. All the coefficients are listed in TABLE I.

### B. FIRST-PRINCIPLES CALCULATION DETAILS

The atomic relaxation and electronic structure calculation were performed by Vienna ab-initio simulation package (VASP).[30, 31] The energy cutoff was chosen as 550 eV and the local density approximation was used with the method of projector augmented-wave.[32] The O 2s2p, Ti 3s3p3d4s, and Pb 5d6s6p electrons are treated as the valence electrons. The optimized lattice constants $a$ and $c$ of PTO are 3.867 and 4.033 Å, respectively, consistent with previous calculation results.[21, 22]

The 180 ° DW models were built by aligning several oppositely-oriented tetragonal unit cells along the $x$ direction. We chose $Nx = 12$, which is large enough to simulate the domain structure of PTO, according to our previous studies.[33] These lattice parameters were fixed during the atomic relaxation to obtain the optimized 180 ° DW models. The k-point mesh was chosen as $1 \times 6 \times 6$. The ionic relaxation was considered as convergent when the Hellmann-Feynman (HF) force on each ion is less than 2 meV/Å.

For a PTO unit cell, its polarization can be calculated by the berry-phase method.[34] For the 180 ° DW model, the polarization of each unit cell can be calculated by the Born effective charge method.[25, 33, 35]

### III. RESULTS AND DISCUSSIONS
### A. THE EFFECT OF FLEXOELECTRIC COEFFICIENTS

For materials with cubic symmetry, there are three non-zero independent flexoelectric coefficients: $f_{1111}$, $f_{1122}$, and $f_{1212}$, which are abbreviated as $f_{11}$, $f_{12}$, and $f_{44}$ using the Voigt notation. We performed several testing calculations by individually setting one of three coefficients as positive and negative values and keeping the other two zero. The results are listed in TABLE II. It is found that when the flexoelectric effect is not taken into account, there are no induced Néel components, and when flexoelectric coefficients change sign, the direction of Néel components also changes. Only $f_{11}$ and $f_{12}$ contribute to the formation of Néel components, while $f_{44}$ has no effects. The polarization distribution around the DW in the case of $f_{12} = -0.1$ (normalized value) is shown in FIG. 1, as an example.

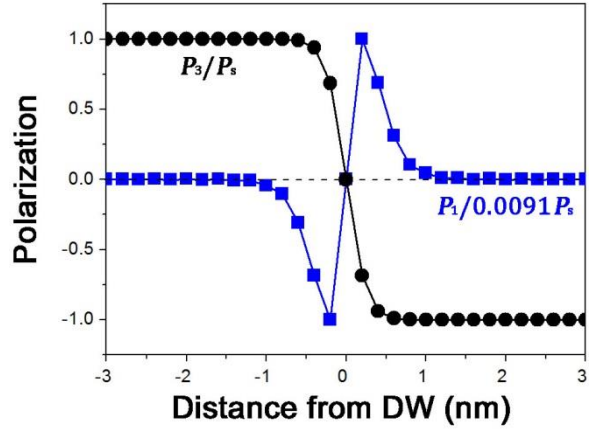

FIG. 1 The distributions of polarization components around the 180° DW in PTO obtained from the phase-field simulations with $f_{12} = -0.1$. $P_s$ is 0.75 C/m$^2$.

The effect of $f_{11}$ and $f_{12}$ can be understood by the flexoelectric field in the $x$ direction:

$$E_1 = f_{1jkl} e_{kl,j} = f_{11} e_{11,1} + f_{12}(e_{22,1} + e_{33,1}) \tag{9}$$

The elastic strain components read

$$e_{11} = \varepsilon_{11} - \varepsilon_{11}^o - \varepsilon_{11}^f = \varepsilon_{11} - Q_{11} P_1^2 - Q_{12} P_2^2 - Q_{12} P_3^2$$
$$+ F_{11} P_{1,1} + F_{12} P_{2,2} + F_{12} P_{3,3} \approx \varepsilon_{11} - Q_{12} P_3^2 \tag{10a}$$

$$e_{22} = \varepsilon_{22} - \varepsilon_{22}^o - \varepsilon_{22}^f \approx \varepsilon_{22} - Q_{12} P_3^2 \tag{10b}$$

$$e_{33} = \varepsilon_{33} - \varepsilon_{33}^o - \varepsilon_{33}^f \approx \varepsilon_{33} - Q_{11} P_3^2 \tag{10c}$$

TABLE II. Maximal Néel-type polarizations and the configuration of Néel components at different combinations of flexoelectric coefficients obtained from phase-field simulations. "T" represents the "tail-to-tail" type and "H" the "head-to-head" type.

| Flexoelectric coefficients (normalized value) | Max. Néel component (unit: $P_s$) | Configuration of Néel components |
| --- | --- | --- |
| $f_{11} = f_{12} = f_{44} = 0$ | 0 | - |
| $f_{11} = +0.1, \ f_{12} = f_{44} = 0$ | $4.1 \times 10^{-3}$ | T |
| $f_{11} = -0.1, \ f_{12} = f_{44} = 0$ | | H |
| $f_{12} = +0.1, \ f_{11} = f_{44} = 0$ | $9.1 \times 10^{-3}$ | H |
| $f_{12} = -0.1, \ f_{11} = f_{44} = 0$ | | T |
| $f_{44} = \pm 0.1, \ f_{11} = f_{12} = 0$ | 0 | - |

We have drawn the distributions of total strains and elastic strains, as presented in FIG. 2. It is found that the total strains $\varepsilon_{22}$ and $\varepsilon_{33}$ are almost constant and the variance of $\varepsilon_{11}$ is also very small. As a contrast, large gradients of elastic strains exist around DWs. Inserting the formulae of elastic strain into Eq. (9), the electric field in the $x$ direction can be written as

$$E_1 \approx -[f_{11}Q_{12} + f_{12}(Q_{11} + Q_{12})](P_3^2)_{,1} \quad (11)$$

The term containing $f_{11}$ could be omitted since $f_{11}$ is usually one order of magnitude smaller than $f_{12}$.[36] For many perovskite oxides, $Q_{11} + Q_{12}$ is positive and $f_{12}$ is negative.[29, 37] As a result, the above formula can be written as

$$E_1 = A(P_3^2)_{,1} \quad (12)$$

where $A$ is a positive number.

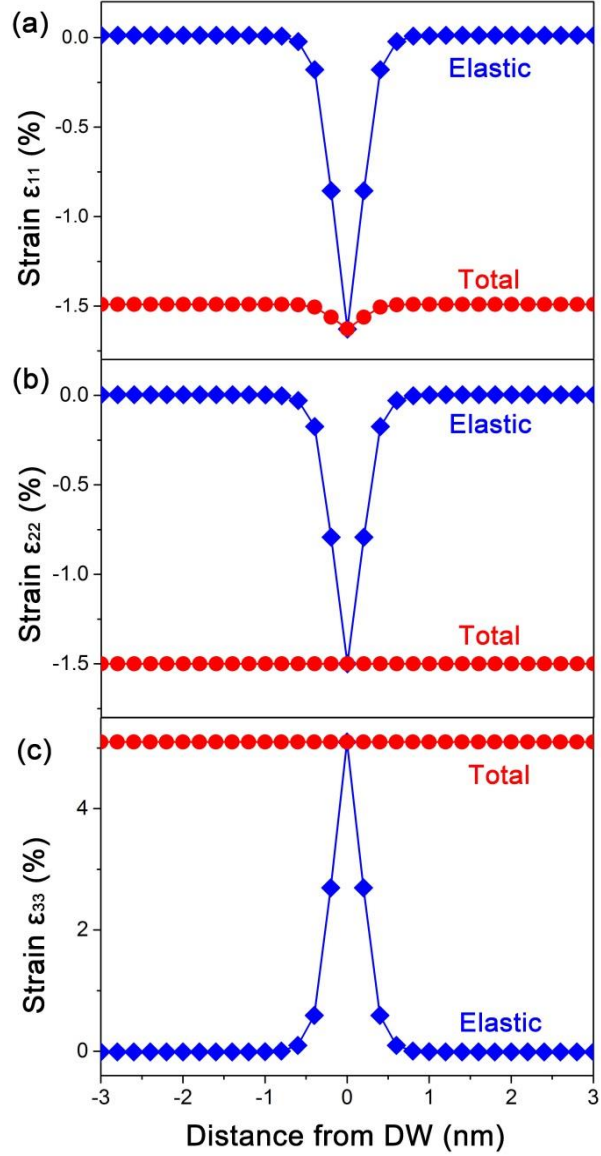

FIG. 2. The elastic and total strain distributions of $\varepsilon_{11}$ (a), $\varepsilon_{22}$ (b), and $\varepsilon_{33}$ (c) around the 180°DW in PTO obtained from the phase-field simulations with $f_{12} = -0.1$.

In FIG. 3, we plotted the flexoelectric field and the depolarization field. It is found that the flexoelectric field forms a tail-to-tail distribution. As a result, the induced Néel components also adopt a tail-to-tail distribution. The depolarization field aroused by Néel components thus forms a head-to-head distribution. The magnitudes of Néel components are the result of the competition between the flexoelectric field and the depolarization field.[28]

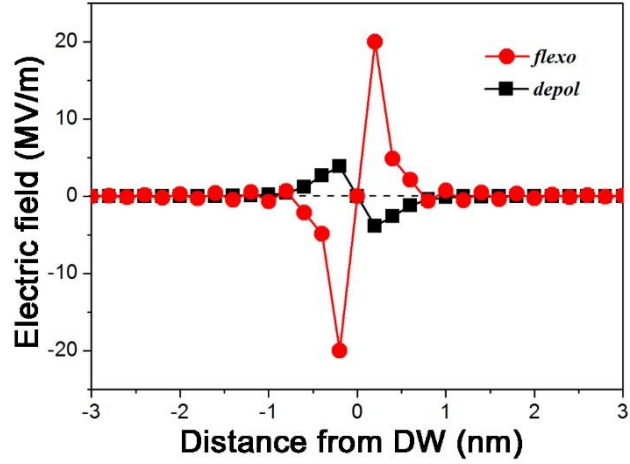

FIG. 3. The distributions of flexoelectric and depolarization fields around the 180° DW in PTO obtained from the phase-field simulations with $f_{12} = -0.1$.

B. ELECTRONIC STRUCTURAL ANALYSIS OF FLEXOELECTRICITY AT 180° DOMAIN WALLS

To understand the flexoelectric effect around 180° DWs more deeply, first-principles calculations were performed. FIG. 4a gives the polarization distributions of Ising and Néel components in the optimized 180° DW model. The tail-to-tail distribution of Néel components is obtained, consistent with the phase-field results and phenomenological analysis. The tail-to-tail distribution of Néel components means that there exists a bound charge around the 180° DW. The bound charge can be calculated according to the formula: $\rho^{bound} = -\nabla \cdot \boldsymbol{P}$, following the method of Li et al.[38] FIG. 4b gives the bound charge distribution of the DW model. It can be found that there exists a negatively charged region at the center of a DW and two positively charged regions nearby. The extreme values of negative and positive charges are about $-3.0 \times 10^7$ C/m$^3$ and $1.6 \times 10^7$ C/m$^3$, respectively. They are about one order of magnitude larger than those around the 180° DW in BaTiO$_3$.[38] The reason may be that PbTiO$_3$ has larger spontaneous polarization and the gradient of the square of spontaneous polarization at 180° DWs is also larger.

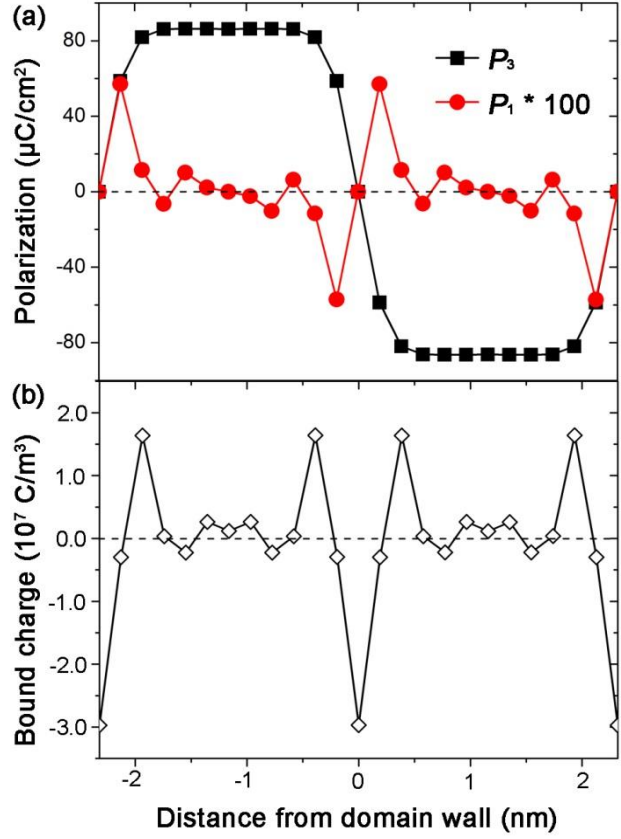

FIG. 4. (a) The distributions of Ising (black blocks) and Néel (red circles) components around 180°DWs in PTO obtained from first-principles calculations. (b) The bound charge distribution calculated from the Néel polarization distribution.

To uncover the ultimate source of Néel components at the electronic level, we artificially removed ferroelectric displacements in the *x* direction (Néel), while kept those in the *z* direction (Ising). Then, the unit-cell-averaged electron charge density and potential distributions along the *x* direction are calculated according to the method used in Ref. 35, as shown in Fig. 5. The distribution of the electric field component in the *x* direction is also shown, by differentiating the potential with the *x* coordinate. From FIG. 5a, we can see that there is an accumulation zone of positive charge localized in about one unit cell at the DW and two negative charged zone at both sides. The magnitudes of negative and positive charges in FIG. 5a are about $-2.0 \times 10^7$ C/m$^3$ and $1.2 \times 10^7$ C/m$^3$, respectively, whose absolute values are in the same order of magnitude as the bound charges in FIG. 4b. Due to the charge separation, a potential difference is built between the DW and the bulk region, as shown in Fig. 5b. This potential difference will generate a tail-to-tail electric field distribution (Fig. 5c) and induce a tail-to-tail $P_1$ distribution. The maximal electric field could reach $3 \times 10^8$ V/m.

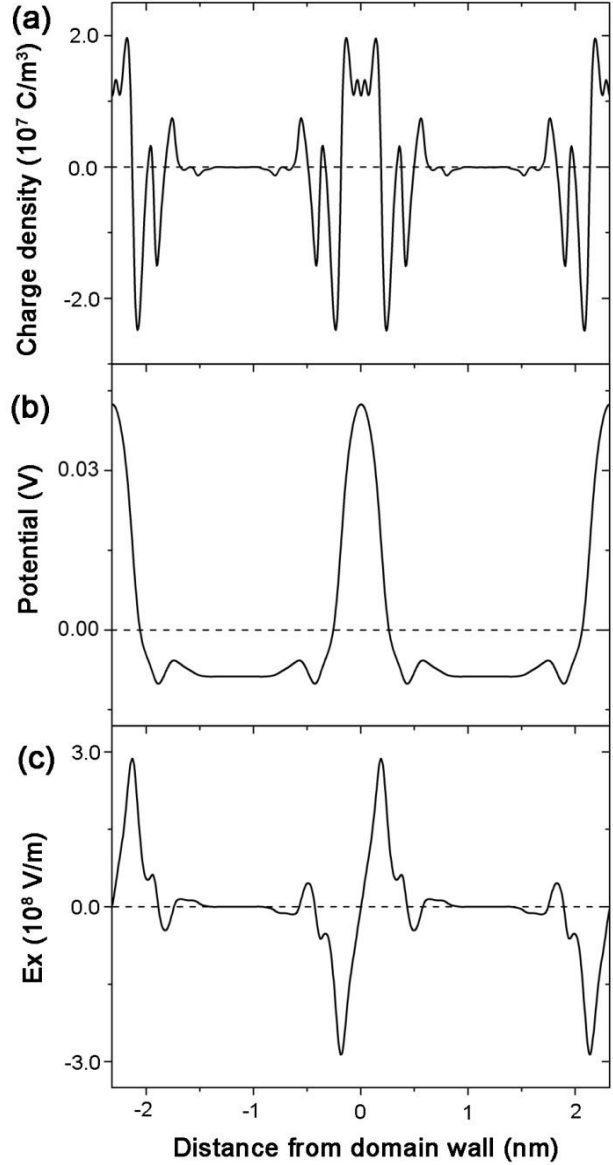

FIG. 5. The distribution of averaged charge density (a), potential (b) and electric field in the $x$ direction (c) around 180° DWs in PTO obtained from first-principles calculations.

The result that there is a positive charge accumulation at Ising DWs seems to contradict with the common knowledge that Ising DWs should be non-conductive. Actually, the charge density is obtained by fixing the atomic coordinates and doing the electronic optimization for the Ising DW model. As a result, there are no ionic polarizations in the *x* direction, while the electronic polarizations exist. Strictly speaking, it cannot be called as the Ising DW model now. Thus, the development of Neel components can be considered as a two-step process: First, an electron redistribution spontaneously occurs in the Ising DW and a potential difference is built between the DW region and the bulk region; Second, ionic displacement normal to the DW is induced by this potential difference, resulting in Néel components.

Comparing FIG. 4b with FIG. 5a, we can find that the two charge distributions show opposite trends. This observation could help us to understand the emergence of Néel components from another aspect: The gradient of the square of Ising components produces a nonuniformly distributed elastic strain (stress). As a result, a nonuniformly distributed charge density forms (FIG. 5a). To compensate this charge density, Néel components develop with oppositely distributed bound charge (FIG. 4b).

## C. THE COMPETITION OF ISING AND BLOCH COMPONENTS ON THE FORMATION OF NÉEL COMPONENTS

For ferroelectric $PbTiO_3$, it is predicted that large Bloch components comparable to Ising ones could develop at 180° DWs, resulting a ferroelectric transition at ferroelectric DWs.[33, 39] These large Bloch components should definitely affect the Néel components through the flexoelectric effect. In FIGs. 4 & 5, we considered the case that only Ising components exist. If Bloch components are taken into account, two other cases come about: The one is that only Bloch components exist and the other is that both Ising and Bloch components exist. The three cases are named as Ising-only, Bloch-only, and Ising-Bloch. The last case corresponds to the real situation.

FIG. 6 gives the distributions of polarizations, charge densities and potentials of the second and third cases. It is found that head-to-head Néel components are induced by the Bloch components, as shown in FIG. 6a. As a contrast, tail-to-tail Néel components are induced by the Ising components, as shown in FIG. 4a. From the charge density and potential profiles of the Bloch-only case shown in FIGs. 6b & 6c, it is found that DWs are the accumulation zones of negative charge and the potential valley. Thus, head-to-head electric field and Néel component distributions are induced. In the Ising-Bloch case, the induced Néel components adopt a tail-to-tail distribution as the Ising-only case (FIG. 6d), which indicates that Ising components are more "powerful" than Bloch components. However, the magnitudes of Néel components are largely reduced (peak values: 0.21 vs 0.58 μC/cm$^2$). Also similar to the Ising-only case, DWs in the Ising-Bloch case are the accumulation zones of positive charge and potential peak (FIGs. 6e & 6f). That is why tail-to-tail Néel component distribution is induced.

We can also understand the cases of Bloch-only and Ising-Bloch phenomenologically. The equations similar to Eq. (12) for the two cases can be written as

$$E_1 \approx A(P_2^2)_{,1} \tag{13a}$$

$$E_1 \approx A(P_2^2 + P_3^2)_{,1} \tag{13b}$$

where the coefficient $A$ is the same as the one in Eq. (12).

The distributions of $P_2^2$, $P_3^2$, $P_2^2 + P_3^2$, and their gradients are shown in FIG. 7. It is found that $P_2^2$ shows peaks at DWs, while both $P_3^2$ and $P_2^2 + P_3^2$ shows valleys at DWs. As a result, the gradient of $P_2^2$ is head-to-head, while the gradients of $P_3^2$ and $P_2^2 + P_3^2$ are tail-to-tail.

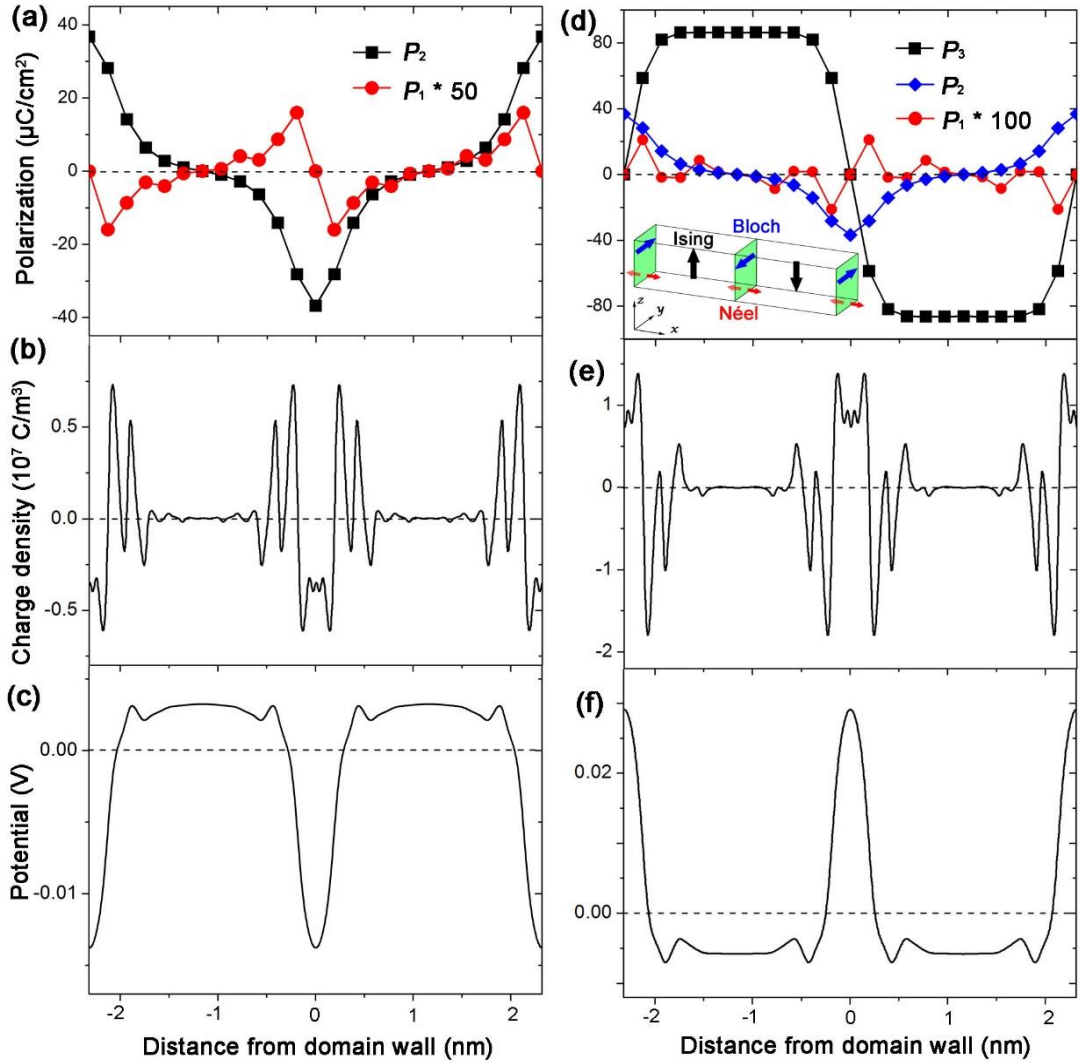

FIG. 6. The distributions of polarization, averaged charge density, and potential for the cases of Bloch-only (left panel), and Ising-Bloch (right panel) around 180° DWs in PTO obtained from first-principles calculations. The distributions of Ising, Bloch, and Néel components in a DW model are schematically shown as an insert in (d).

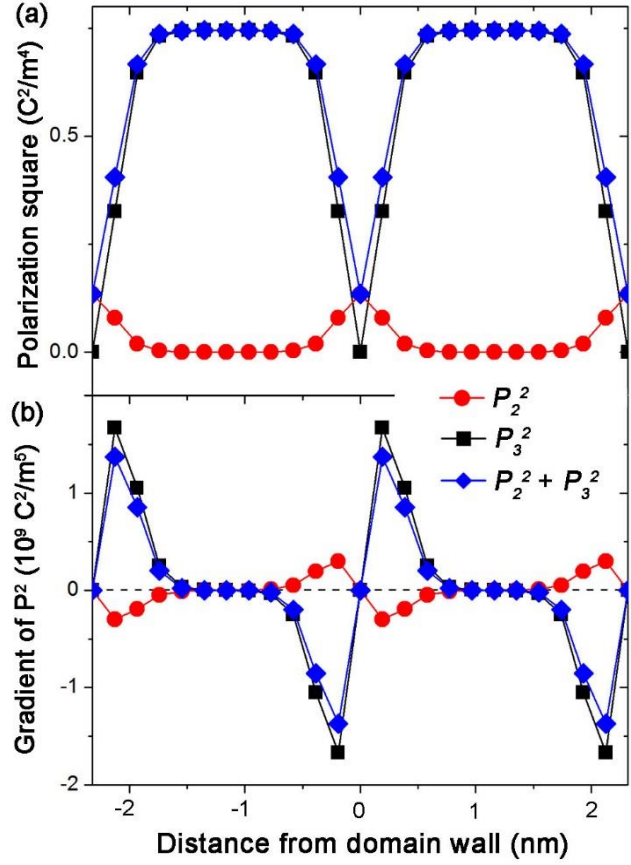

FIG. 7. The distributions of $P_2^2$, $P_3^2$, $P_2^2 + P_3^2$ (a), and their gradients (b) around 180° DWs in PTO obtained from first-principles calculations.

## IV. CONCLUSIONS

In this work, we used phase-field simulations to study the effect of different flexoelectric coefficients on the distribution of Néel components around 180° DWs in tetragonal ferroelectric PTO and further used first-principles calculations to explore the electronic origin of the flexoelectric effect. The main conclusions are listed as follows:

1. The driving force of Néel components comes from the flexoelectric coefficient $f_{11}$ and $f_{12}$ and the ultimate source of Néel components is the gradient of the square of spontaneous polarizations.

2. Electronic structural analysis reveals that there is an accumulation zone of positive charge around the Ising type 180° DWs, which results in the potential difference, tail-to-tail electric field and Néel components.

3. The contributions of Ising and Bloch components on the formation of Néel components are opposite and the competition result is a tail-to-tail distribution of Néel components with reduced magnitudes.

**Acknowledgments:** This work is supported by National Key Basic Research Program of China (2016YFA0201001, 2014CB921002), National Natural Science Foundation of China (No. 51401212, 11627801, 51571197, 51231007, 51671194, and 51521091), Key Research Program of Frontier Sciences CAS (QYZDJ-SSW-JSC010), and Doctoral Initiation Foundation of Liaoning Province (No. 20141144).

[a) b)] Author to whom correspondence should be addressed.
Electronic mail: xlma@imr.ac.cn, jjli@uw.edu

# Supplementary materials to

# "Phase-field modeling and electronic structural analysis of flexoelectric effect at 180 ° domain walls in ferroelectric PbTiO$_3$"

Yu-Jia Wang[1], Jiangyu Li[2,3,a)], Yin-Lian Zhu[1], Xiu-Liang Ma[1,4,b)]

[1]Shenyang National Laboratory for Materials Science, Institute of Metal Research, Chinese Academy of Sciences, 72 Wenhua Road, Shenyang, Liaoning, 110016, China

[2]Shenzhen Key Laboratory of Nanobiomechanics, Shenzhen Institutes of Advanced Technology, Chinese Academy of Sciences, University Town of Shenzhen, Shenzhen, Guangdong, 518055, China

[3]Department of Mechanical Engineering, University of Washington, Seattle, Washington 98195-2600, USA

[4]School of Materials Science and Engineering, Lanzhou University of Technology, Langongping Road 287, 730050 Lanzhou, China

## THE INCORPORATING OF THE FLEXOELECTRIC EFFECT INTO THE ELASTIC ENERGY

In Ref. 1, the phenomenological theory of the flexoelectric effect at the stress-free boundary condition is established. The elastic Gibbs free energy is chosen as the system's thermodynamic potential, whose variables are the polarization and stress. It is alternative to use the Helmholtz free energy whose variables are the polarization and strain, by doing the Legendre transformation. The stress-related terms in the elastic Gibbs free energy in Ref. 1 can be written as

$$G = -\frac{1}{2}s_{ijkl}\sigma_{ij}\sigma_{kl} - Q_{ijkl}\sigma_{ij}P_k P_l + F_{ijkl}\sigma_{ij}P_{k,l} \qquad (S1)$$

where, $s_{ijkl}$ is the elastic compliance. In other literatures[2, 3], the flexoelectric coupling term is often written as

$$F_{flex} = \frac{1}{2}F'_{ijkl}(\sigma_{ij}P_{k,l} - P_k\sigma_{ij,l}). \tag{S2}$$

As pointed out by Gu[4], these two expressions are equivalent in the condition of $F_{ijkl} = F'_{ijkl}$.

By functional derivative of $G$ with respective to $\sigma_{ij}$, we can get

$$\varepsilon_{ij} = -\frac{\delta G}{\delta \sigma_{ij}} = s_{ijkl}\sigma_{kl} + Q_{ijkl}P_kP_l - F_{ijkl}P_{k,l} \tag{S3a}$$

$$\sigma_{ij} = C_{ijkl}(\varepsilon_{kl} - Q_{klmn}P_mP_n + F_{klmn}P_{m,n}) \tag{S3b}$$

Inserting the expression of $\sigma_{ij}$ into $F = G + \sigma_{ij}\varepsilon_{ij}$, the Helmholtz free energy can be obtained:

$$\begin{aligned}
F &= G + \sigma_{ij}\varepsilon_{ij} \\
&= -\frac{1}{2}C_{ijkl}(\varepsilon_{ij}\varepsilon_{kl} + Q_{ijmn}P_mP_nQ_{klor}P_oP_r - 2\varepsilon_{ij}Q_{klor}P_oP_r \\
&\quad + 2\varepsilon_{ij}F_{klor}P_{o,r} + F_{ijmn}P_{m,n}F_{klor}P_{o,r} - 2Q_{ijmn}P_mP_nF_{klor}P_{o,r}) \\
&\quad - C_{ijkl}(\varepsilon_{ij}Q_{klor}P_oP_r - Q_{ijmn}P_mP_nQ_{klor}P_oP_r + Q_{ijmn}P_mP_nF_{klor}P_{o,r}) \\
&\quad + C_{ijkl}(\varepsilon_{ij}F_{klor}P_{o,r} - Q_{ijmn}P_mP_nF_{klor}P_{o,r} + F_{ijmn}P_{m,n}F_{klor}P_{o,r}) \\
&\quad + C_{ijkl}(\varepsilon_{ij}\varepsilon_{kl} - \varepsilon_{ij}Q_{klor}P_oP_r + \varepsilon_{ij}F_{klor}P_{o,r}) \\
&= \frac{1}{2}C_{ijkl}(\varepsilon_{ij}\varepsilon_{kl} + Q_{ijmn}P_mP_nQ_{klor}P_oP_r - 2\varepsilon_{ij}Q_{klor}P_oP_r \\
&\quad + 2\varepsilon_{ij}F_{klor}P_{o,r} + F_{ijmn}P_{m,n}F_{klor}P_{o,r} - 2Q_{ijmn}P_mP_nF_{klor}P_{o,r}) \\
&= \frac{1}{2}C_{ijkl}(\varepsilon_{ij} - \varepsilon_{ij}^o - \varepsilon_{ij}^f)(\varepsilon_{kl} - \varepsilon_{kl}^o - \varepsilon_{kl}^f)
\end{aligned}$$

$$\tag{S4}$$

This expression is what we used in our phase-field model, which contains six terms. The first term is the elastic energy which comes from the total strain:

$$F_{E1} = \frac{1}{2}C_{ijkl}\varepsilon_{ij}\varepsilon_{kl}$$

$$= \frac{1}{2}C_{11}(\varepsilon_{11}^2 + \varepsilon_{22}^2 + \varepsilon_{33}^2) + C_{12}(\varepsilon_{11}\varepsilon_{22} + \varepsilon_{11}\varepsilon_{33} + \varepsilon_{22}\varepsilon_{33})$$

$$+ 2C_{44}(\varepsilon_{12}^2 + \varepsilon_{13}^2 + \varepsilon_{23}^2)$$

(S5)

The second term is the modification of the fourth order terms in the Landau-Devonshire energy:

$$F_{E2} = \frac{1}{2}C_{ijkl}Q_{ijmn}P_m P_n Q_{klop}P_o P_p$$

$$= \beta_{11}(P_1^4 + P_2^4 + P_3^4) + \beta_{12}(P_1^2 P_2^2 + P_1^2 P_3^2 + P_2^2 P_3^2)$$

(S6a)

where

$$\beta_{11} = \frac{1}{2}C_{11}(Q_{11}^2 + 2Q_{12}^2) + C_{12}(2Q_{11}Q_{12} + Q_{12}^2) \quad \text{(S6b)}$$

$$\beta_{12} = C_{12}(2Q_{11}Q_{12} + Q_{12}^2) + C_{12}(Q_{11}^2 + 3Q_{12}^2 + 2Q_{11}Q_{12}) + 2C_{44}Q_{44}^2 \quad \text{(S6c)}$$

The third term is the electrostrictive coupling between the total strain and the polarization:

$$F_{E3} = -C_{ijkl}\varepsilon_{ij}Q_{klop}P_o P_p$$

$$= -(q_{11}\varepsilon_{11} + q_{12}\varepsilon_{22} + q_{12}\varepsilon_{33})P_1^2 - (q_{12}\varepsilon_{11} + q_{11}\varepsilon_{22} + q_{12}\varepsilon_{33})P_2^2$$

$$- (q_{12}\varepsilon_{11} + q_{12}\varepsilon_{22} + q_{11}\varepsilon_{33})P_2^2$$

$$- 2q_{44}(\varepsilon_{12}P_1 P_2 + \varepsilon_{23}P_2 P_3 + \varepsilon_{13}P_1 P_3)$$

(S7a)

where

$$q_{11} = C_{11}Q_{11} + 2C_{12}Q_{12} \quad \text{(S7b)}$$

$$q_{12} = C_{11}Q_{12} + C_{12}Q_{11} + C_{12}Q_{12} \quad \text{(S7c)}$$

$$q_{44} = 2C_{44}Q_{44} \quad \text{(S7d)}$$

The fourth term is the flexoelectric coupling between the total strain and the gradient of polarization, which can be equivalently written as $\frac{1}{2}f_{ijkl}(\varepsilon_{ij}P_{k,l} - P_k \varepsilon_{ij,l})$.

$$F_{E4} = C_{ijkl}\varepsilon_{ij}F_{klop}P_{o,p}$$
$$= (f_{11}\varepsilon_{11} + f_{12}\varepsilon_{22} + f_{12}\varepsilon_{33})P_{1,1} + (f_{12}\varepsilon_{11} + f_{11}\varepsilon_{22} + f_{12}\varepsilon_{33})P_{2,2}$$
$$+ (f_{12}\varepsilon_{11} + f_{12}\varepsilon_{22} + f_{11}\varepsilon_{33})P_{3,3}$$
$$+ 2f_{44}[\varepsilon_{12}(P_{1,2} + P_{2,1}) + \varepsilon_{13}(P_{1,3} + P_{3,1}) + \varepsilon_{23}(P_{2,3} + P_{3,2})]$$

(S8a)

where

$$f_{11} = C_{11}F_{11} + 2C_{12}F_{12} \quad \text{(S8b)}$$
$$f_{12} = C_{11}F_{12} + C_{12}F_{11} + C_{12}F_{12} \quad \text{(S8c)}$$
$$f_{44} = 2C_{44}F_{44} \quad \text{(S8d)}$$

The fifth term is the square of the polarization gradient, whose effect is the renormalization of the gradient energy.

$$F_{E5} = \frac{1}{2}C_{ijkl}F_{ijmn}P_{m,n}F_{klop}P_{o,p}$$
$$= \frac{1}{2}G_{11}^{f}\left[(P_{1,1})^2 + (P_{2,2})^2 + (P_{3,3})^2\right]$$
$$+ G_{12}^{f}(P_{1,1}P_{2,2} + P_{1,1}P_{3,3} + P_{2,2}P_{3,3})$$
$$+ \frac{1}{2}G_{44}^{f}\left[(P_{1,2} + P_{2,1})^2 + (P_{1,3} + P_{3,1})^2 + (P_{2,3} + P_{3,2})^2\right]$$

(S9a)

where

$$G_{11}^{f} = C_{11}(F_{11}^2 + 2F_{12}^2) + 2C_{12}(2F_{11}F_{12} + F_{12}^2) \quad \text{(S9b)}$$
$$G_{12}^{f} = C_{11}(2F_{11}F_{12} + F_{12}^2) + C_{12}(F_{11}^2 + 3F_{12}^2 + 2F_{11}F_{12}) \quad \text{(S9c)}$$
$$G_{44}^{f} = 4C_{44}F_{44}^2 \quad \text{(S9d)}$$

The sixth term is the coupling between the electrostrictive strain and the flexoelectric strain. This term is most important to obtain the correct driving force of Neel components.

$$F_{E6} = -C_{ijkl}Q_{ijmn}P_m P_n F_{klop}P_{o,p}$$
$$= -(\gamma_{11}P_1^2 + \gamma_{12}P_2^2 + \gamma_{12}P_3^2)P_{1,1} - (\gamma_{12}P_1^2 + \gamma_{11}P_2^2 + \gamma_{12}P_3^2)P_{2,2}$$
$$- (\gamma_{12}P_1^2 + \gamma_{12}P_2^2 + \gamma_{11}P_3^2)P_{3,3}$$
$$- 2\gamma_{44}[P_1P_2(P_{1,2} + P_{2,1}) + P_1P_3(P_{1,3} + P_{3,1}) + P_2P_3(P_{2,3} + P_{3,2})]$$

(S10a)

where

$$\gamma_{11} = C_{11}(Q_{11}F_{11} + 2Q_{12}F_{12}) + 2C_{12}(Q_{11}F_{12} + Q_{12}F_{11} + Q_{12}F_{12}) \quad \text{(S10b)}$$

$$\gamma_{12} = C_{11}(Q_{11}F_{12} + Q_{12}F_{11} + Q_{12}F_{12})$$
$$+ C_{12}(Q_{11}F_{11} + Q_{11}F_{12} + Q_{12}F_{11} + 3Q_{12}F_{12}) \quad \text{(S10c)}$$

$$\gamma_{44} = 2C_{44}Q_{44}F_{44} \quad \text{(S10d)}$$

The mechanical driving force can be found by the functional derivative of $F$ with respective to $P$:

$$-\frac{\delta F_{elas}}{\delta P_i} = -\left[\frac{\partial F_{elas}}{\partial P_i} - \frac{\partial}{\partial x_j}\left(\frac{\partial F_{elas}}{\partial P_{i,j}}\right)\right] = -\frac{\partial F_{elas}}{\partial P_i} + \frac{\partial}{\partial x_r}\left(\frac{\partial F_{elas}}{\partial P_{i,r}}\right)$$

$$= -\frac{\partial F_{elas}}{\partial P_i} + \frac{\partial}{\partial x_r}\left[\left(\frac{\partial F_{elas}}{\partial P_{o,r}}\right)\frac{\delta P_{o,r}}{\delta P_{i,r}}\right] = -\frac{\partial F_{elas}}{\partial P_i} + \delta_{oi}\frac{\partial}{\partial x_r}\left(\frac{\partial F_{elas}}{\partial P_{o,r}}\right)$$

(S11a)

$$-\frac{\partial F_{elas}}{\partial P_i} = -\frac{1}{2}C_{ijkl}\big[Q_{ijmn}Q_{klor}(\delta_{mi}P_nP_oP_r + \delta_{ni}P_mP_oP_r + \delta_{oi}P_mP_nP_r + \delta_{ri}P_mP_nP_o)$$
$$- 2Q_{klor}\varepsilon_{ij}(\delta_{oi}P_r + \delta_{ri}P_o) - 2Q_{ijmn}F_{klor}(\delta_{mi}P_n + \delta_{ni}P_m)P_{o,r}\big]$$

$$= -\frac{1}{2}C_{ijkl}\big(4Q_{ijin}Q_{klor}P_nP_oP_r - 4Q_{ijin}\varepsilon_{kl}P_n - 4Q_{ijin}F_{klor}P_nP_{o,r}\big)$$

$$= 2C_{ijkl}Q_{ijin}P_n(\varepsilon_{kl} - Q_{klor}P_oP_r + F_{klor}P_{o,r}) = 2C_{ijkl}Q_{ijin}P_ne_{kl}$$

$$= 2q_{inkl}P_ne_{kl} = 2q_{ijkl}P_je_{kl}$$

(S11b)

$$\delta_{oi}\frac{\partial}{\partial x_r}\left(\frac{\partial F_{elas}}{\partial P_{o,r}}\right) = \frac{1}{2}C_{ijkl}\delta_{oi}\frac{\partial}{\partial x_r}(2F_{klor}\varepsilon_{ij} + 2F_{ijmn}F_{klor}P_{m,n} - 2Q_{ijmn}F_{klor}P_mP_n)$$

$$= \frac{1}{2}C_{ijkl}\big[2F_{klir}\varepsilon_{ij,r} + 2F_{klir}F_{ijmn}P_{m,nr} - 2F_{klir}Q_{ijmn}(P_mP_n)_{,r}\big]$$

$$= C_{ijkl}F_{klir}\big[\varepsilon_{ij,r} - Q_{ijmn}(P_mP_n)_{,r} + F_{ijmn}P_{m,nr}\big] = C_{ijkl}F_{klir}e_{ij,r}$$

$$= C_{ijkl}F_{ijir}e_{kl,r} = f_{irkl}e_{kl,r} = f_{ijkl}e_{kl,j}$$

(S11c)

As a result, the mechanical driving force is $2q_{ijkl}P_j e_{kl} + f_{ijkl}e_{kl,j}$. The first part contains the elastic strain, while the second part contains the gradient of the elastic strain. Flexoelectric effect is embedded in both parts. One term in the second part is $-C_{ijkl}F_{klir}Q_{ijmn}(P_m P_n)_{,r} = -f_{ijkl}Q_{klmn}(P_m P_n)_{,j}$, which comes from the sixth term in Eq. 1 and is the main driving force of Neel components: $E_1 \approx -[f_{11}Q_{12} + f_{12}(Q_{11} + Q_{12})](P_3^2)_{,1}$.